\documentclass[preprintnumbers,amsmath,amssymb]{revtex4}
\usepackage{graphics}
\usepackage{dcolumn}

\begin{document}

\title{Self-interacting scalar field in (2+1) dimensions Einstein gravity with torsion }
\author{R. Kaya*, H. T. \"{O}z\c{c}elik*}
\address{*Yildiz Technical University, Department of Physics, 34220 Davutpa\c sa, Istanbul, TURKEY}

\begin{abstract}
We study a massless real self-interacting scalar field $\varphi$
non-minimally coupled to Einstein gravity with torsion in (2+1)
space-time dimensions in the presence of cosmological constant.
The field equations with a self-interaction potential $V(\varphi)$
including $\varphi^{n}$ terms are derived by a variational
principle. By numerically solving these field equations with the
4th Runge-Kutta method, the circularly symmetric rotating
solutions for (2+1) dimensions Einstein gravity with torsion are
obtained. Exact analytical solutions to the field equations are
derived for the proposed metric in the absence of both torsion and
angular momentum. We find that the self-interacting potential only
exists for $n=6$. We also study the motion of massive and massless
particles in (2+1) Einstein gravity with torsion coupled to a
self-interacting scalar field. The effect of torsion on the
behavior of the effective potentials of the particles is analyzed
numerically.
\end{abstract}

\keywords{Torsion\sep  Self-interacting \sep  Scalar field \sep
Particle motion}

\maketitle

\section{\label{sec:level1}Introduction}

In recent years there has been much more interest in the self-interacting scalar field, which has played a great role in general relativity, astrophysics, cosmology, and unified theories of elementary particles \cite{avr,olaf,dzh,dmit,sergey,andre,char,alej,sour,Linde,Ghilenca}. These fields in the framework of general relativity are often used to describe the behavior of the early universe, the inflationary period \cite{bru,gut}, dark energy which could explain the accelerated expansion of the universe \cite{riess,perl}, dark matter \cite{ton1,ton2}, and other cosmological phenomena.

The self-interacting field theory is more complex than the free scalar field
theory. As the self-interacting field theory includes the interaction terms, it is
more difficult to find the solution of Einstein and Klein-Gordon field equations.

The solutions of Einstein gravity for minimally and non-minimally coupled scalar fields without the self-interaction
term have been obtained \cite{19,20,Ozcelik1,ayse,21,hasn,sing,halil,mar2}.

In the past years, a potential explanation of scalar field theories has been given by a generalization of Einstein gravity with torsion, where the torsion field is generated by a scalar field. Torsion has important consequences from the phenomenological point of view. In this respect, the non-minimally coupled scalar field with curvature and torsion in four and three dimensions has been studied in the literature \cite{sur1,sur2,sur3,gal1,gal2,gal3,gal4,gal5,cid,Ozcelik2}.

In this paper, we consider a massless self-interacting scalar field, non-minimally coupled to Einstein gravity with torsion in (2+1) space-time dimensions, in the presence of a cosmological constant $\Lambda$. Firstly, we define the action including the self-interacting potential. Then, we obtain Einstein field equations and the Klein-Gordon equation derived from the variation of the action with respect to the dreibein field and the scalar field. We find the exact solution of them when the torsion and angular momentum vanish. We could not solve field equations in the presence of both torsion and angular momentum. These equations are solved numerically with the 4th Runge-Kutta method.

To discover the space-time structure, the motion of test particles can be used to classify an arbitrary space-time.
The stability of circular orbits of particles has been studied in the four-dimensional gravity. Moreover, these studies have been extended to the five-dimensional gravity. The motion of test particles in (2+1) dimensional gravity with and without torsion has been also investigated in Refs. \cite{far,sha2,sah,gha,gon,hend,sob,Kaya1}.

We also investigate torsion effects on massive and massless particles in the (2+1) dimensions of Einstein gravity. We derive the equations of motion for massive and massless particles by using the Euler-Lagrange equation. We give the effective potential for the radial motion for self-interacting scalar fields in this space-time.

The outline of this paper is as follows. In Sec. 2 we introduce the metric and obtain the Einstein field equations and Klein-Gordon equation. Sec. 3 includes a class of solutions to these field equations. In Sec. 4, we present properties of the motion of massive and massless particles in the (2+1) dimensions of Einstein gravity with and without torsion. Finally, our conclusions are given in Sec. 5.

\section{\label{sec:level1}Einstein gravity with torsion induced by the self-interacting scalar field}

We consider the massless scalar field interacting with itself as the source of torsion. The action for a massless self-interacting scalar field non-minimally conformal coupled to Einstein gravity with torsion in (2+1) space-time dimensions in the presence of a cosmological constant $\Lambda$ is given by
\begin{eqnarray}
S&=
&\int\sqrt{-g}\;d^3x \;   \big(\frac{1}{2\kappa}(R-2\Lambda)-\frac{1}{2}
\nabla^{\mu}\varphi\nabla_{\mu}\varphi \nonumber\\ &&\hspace{5mm}-\frac{1}{2} \xi R\varphi^2 +V(\varphi) \big)
\end{eqnarray}
where $\kappa=8 \pi G /c^4$ is the Einstein gravitational constant, $R=g^{\mu\nu}R_{\mu \nu}$ is the curvature scalar of the Riemann-Cartan space-time, $g$ is the metric tensor determinant, $\varphi$ is the scalar field and $V(\varphi)$ is a potential of self-interaction of the scalar field
\begin{eqnarray}
V(\varphi)=\sum _{n=1} {\lambda _n}{\varphi ^n},
\end{eqnarray}
with the coupling constant $ \lambda_n $ describing the strength of the interaction between the scalar fields.

The non-minimally coupling constant $\xi$ will be taken to be equal to $\frac{1}{8}$ in (2+1) dimensions, resulting in a conformally invariant field theory \cite{birell}.

The Riemann-Cartan curvature tensor $R^\rho_{\sigma\mu\nu}$ and Ricci tensor $ R_{\mu\nu}$ are defined as
\begin{eqnarray}
R^\rho_{\sigma\mu\nu}=\partial_\mu\Gamma^\rho_{\nu\sigma}-\partial_\nu\Gamma^\rho_{\mu\sigma}+\Gamma^\rho_{\mu\delta}
    \Gamma^\delta_{\nu\sigma}
    -\Gamma^\rho_{\nu\delta}\Gamma^\delta_{\mu\sigma},
\end{eqnarray}
and $ R_{\mu\nu} = R_{\mu \rho \nu}^{\rho} $.

The anti-symmetric part of the affine connection $\Gamma_{\mu\nu }^{\rho }$ defines the non-zero torsion tensor
 $ {T_{\mu\nu}}^\rho=\frac{1}{2}(\Gamma^\rho_{\mu\nu}-\Gamma^\rho_{\nu\mu}) $.
The metric tensor $ g_ {\mu\nu} $ can be expressed as
\begin{eqnarray}
g_ {\mu\nu}=\eta_{ab}e_{\mu}{}^{a } e_{\nu}{}^{b },
\end{eqnarray}
where $ \eta_{ab} $ = (--,+,+) is the Minkowski metric tensor, $ e_{\mu}^{\;\;a} $ is the dreibein field.

The Greek indices $=$ 1,\,2,\,3 denote (2+1) space-time coordinates and Roman indices $=$ 1,\,2,\,3 denote locally flat coordinates. Throughout the paper, we use $8 \pi G=\hbar=c=1$.

From the dreibein postulate $\nabla_\mu e_{\nu}^{\;\;a}=0$ one can  obtain the $\Gamma$-connection as follows
\begin{eqnarray}
\Gamma _{\mu  \nu}{}^{\rho }=\{_{\mu  \nu}{}^{\rho } \}+\alpha \, e_a{}^{\rho }\, e_\nu{}^{b }\,\omega _{\mu }{}^{b}{}_a,
\end{eqnarray}
with the zero-torsion part
\begin{eqnarray}
\left.\{_{\mu  v}{}^{\rho }\right\}=e_a{}^{\rho } (\partial_\mu e_\nu {}^a+e_\nu{}^b\, \overset{o}{\omega }_{\mu }{}^b{}_a)
\end{eqnarray}
and  the Lorentz connection field
\begin{eqnarray}
\omega _{\mu }{}^{b}{}_a= \eta_{ca}\,\omega _{\mu }{}^{bc}.
\end{eqnarray}
The second term in Eq. (5) gives the contribution from the torsion. When $\alpha=0$ it means zero-torsion. When $\alpha=1$ it means that there is a torsion.

We take the following circular stationary and rotational symmetric (2+1) dimensional space-time as
\begin{eqnarray}
ds^2 = -(v +\frac{J^{2}}{r^2})dt^2+w^2 dr^2+(r
d\phi+\frac{J}{r}dt)^2.
\end{eqnarray}
Here we assume that the metric components  $v$, $w$ and the angular momentum $ J $ will be functions of the radial coordinate $r$.

An orthonormal base for the metric (8) can be found
\begin{eqnarray}
e_{\mu}{}^a=
\begin{pmatrix}
\frac{\sqrt{J^2 + r^2 v}}{r}\; & 0 \;& 0  \;\\ 0&  \;w \; &\; 0  \;\\  \frac{J}{r}& \; 0 \;& \; r \;
\end{pmatrix}.
\end{eqnarray}
From the metricity condition $\nabla_ \sigma \,g_{\mu \nu}=0$ and $\Gamma$-connection (5), the non-zero components of the Lorentz connection coefficients can be obtained as follows
\begin{eqnarray}
&&\omega_1^{\;23}=-\omega_1^{\;32}= \frac{ J U}{rw}, \hspace{2.5mm}\omega_3^{\;23}=-\omega_3^{\;32} = \frac{rU}{w},\nonumber\\&&\omega_1^{\;21}=-\omega_1^{\;12}=\frac{U\sqrt{J^2 + r^2 v}}{rw},
\end{eqnarray}
with
\begin{eqnarray}
U=\frac{16 \varphi\varphi'}{\varphi^2-8\,}.
\end{eqnarray}
Here $'$ denotes the derivative with respect to $r$.

\subsection{The field equations}

The variation of the total action (1) with respect to the dreibein field $e_{a}{}^\mu$
\begin{eqnarray}
\frac{\partial (\sqrt{-g}L)}{\partial
e_{a}{}^\mu}-\partial_{\rho}\frac{\partial
(\sqrt{-g}L)}{\partial(\partial_{\rho}e_{a}{}^\mu)}+\partial_{\sigma}\partial_{\rho}\frac{\partial(\sqrt{-g}
L)}{\partial(\partial_{\sigma}\partial_{\rho}e_{a}{}^\mu)}=0,
\end{eqnarray}
yields the Einstein field equations as follows
\begin{eqnarray}
& & \sum _{n=1}^m \lambda_n \varphi ^n-\alpha\frac{  \varphi ^2
(\varphi ')^2}{2 (\varphi ^2-8) w^2}-\frac{1}{32 w^3 (J^2+r^2
v)^2}\big[32 w (\Lambda  w^2 (J^2+r^2 v)^2+J^2 v)+2 (J^2+r^2 v)
(2 w' (2 \varphi  \varphi ' (J^2+r^2 v)\nonumber\\
& &+r v (\varphi ^2-8)) +w (-4 \varphi  \varphi '' (J^2+r^2 v)+4
(\varphi ')^2 (J^2+r^2 v)
-J (\varphi ^2-8)J''-4 r v \varphi  \varphi '))\nonumber\\
& &-2 J^2 r (\varphi ^2-8) w v'-4 J^2 v \varphi ^2 w+J J' ((\varphi ^2-8) (2 w' (J^2+r^2 v)+r w (r v'+6 v))-4 \varphi  w \varphi ' (J^2+r^2 v))\nonumber\\
& &+ (\varphi ^2-8)w (J')^2 (J^2-r^2 v)\big]=0,
\end{eqnarray}
\begin{eqnarray}
& & \sum _{n=1}^m \lambda_n \varphi ^n+\alpha\frac{  \varphi ^2  (\varphi ')^2}{2 (\varphi ^2-8) w^2}+\frac{1}{32 w^3 (J^2+r^2 v)^2}\big[32 w (\Lambda  w^2 (J^2+r^2 v)^2+J^2 v)+2 (J^2+r^2 v)\nonumber\\
& & (2 w' (2 \varphi  \varphi ' (J^2+r^2 v)+r v (\varphi ^2-8)) +w (-4 \varphi  \varphi '' (J^2+r^2 v)+4 (\varphi ')^2 (J^2+r^2 v) \nonumber\\
& &-J (\varphi ^2-8)J''-4 r v \varphi  \varphi '))-2 J^2 r (\varphi ^2-8) w v'-4 J^2 v \varphi ^2 w+J J' ((\varphi ^2-8) (2 w' (J^2+r^2 v)\nonumber\\
& &+r w (r v'+6 v))-4 \varphi  w \varphi ' (J^2+r^2 v))+ (\varphi
^2-8)w (J')^2 (J^2-r^2 v)\big]=0,
\end{eqnarray}
\begin{eqnarray}
& & \sum _{n=1}^m \! \lambda _n \varphi ^n\!-\alpha  \frac{\varphi
^2 (\varphi ')^2}{2 (\varphi ^2-8) w^2}+ \frac{1}{32 r^2 w^3
(J^2+r^2 v)^2}
\big[-2 r (J^2+r^2 v) (w (-r (\varphi ^2-8) (2 J J''+r^2 v'')\nonumber\\
& &-4 (r (J^2+r^2 v)( \varphi  \varphi '' +  (\varphi ')^2 )+ J^2 \varphi  \varphi ')) +2 w'(2 r \varphi  \varphi ' (J^2+r^2 v)-J^2 (\varphi ^2-8)))+2 r^3 v' (r \nonumber\\
& &(J^2+r^2 v) (2 \varphi  w \varphi '-(\varphi ^2-8) w')+ (\varphi ^2-8) 2 J^2 w)-4 w(J^4 (8 \Lambda  r^2 w^2+\varphi ^2-8)+16 J^2 \Lambda  r^4 \nonumber\\
& &v w^2+8 \Lambda  r^6 v^2 w^2)+(\varphi ^2-8) (J')^2 (r^2 v-3 J^2)r^2 w +4 J r J' (2 r \varphi  w \varphi ' (J^2+r^2 v)-(\varphi ^2-8)  \nonumber\\
& &(r w'(J^2+r^2 v)-J^2 w+r^3 w v'+r^2 v w))+r^6 (\varphi ^2-8)
(-w) (v')^2\big]=0,
\end{eqnarray}
\begin{eqnarray}
& & \sum _{n=1}^m \lambda _n \varphi ^n-\alpha \frac{ \varphi ^2
(\varphi ')^2}{2 (\varphi ^2-8) w^2} +\frac{1}{32 J w^3 (J^2+r^2
v)^2} [2 J r v' (r (J^2+r^2 v) (2 \varphi  w \varphi '-(\varphi
^2-8) w')+\nonumber\\& &(\varphi ^2-8) w (J^2-r^2 v))+4 J  (-8
\Lambda  w^2 (J^2+r^2 v)^2+J^2 v \varphi ^2-8 J^2 v)w-2 (J^2+r^2
v) (2 J w' (2 \varphi  \varphi ' \nonumber\\& &(J^2+r^2 v)+r v
(\varphi ^2-8))+w ( (\varphi ')^2 (J^3+J r^2 v)4-(\varphi ^2-8)
(J'' (J^2-r^2 v)+J r^2 v'')-4 J \varphi  \varphi '' \nonumber\\&
&(J^2+r^2 v)-4 J r v \varphi \varphi '))-J (\varphi ^2-8) w (J')^2
(J^2-3 r^2 v)+J' (4 \varphi w \varphi ' (J^4-r^4 v^2)+(\varphi
^2-8)\nonumber\\& & (r w (r^2 v-3 J^2) (r v'+2 v)-2 w' (J^4-r^4
v^2)))-J r^4 (\varphi ^2-8) w (v')^2]=0,
\end{eqnarray}
\begin{eqnarray}
& &4 J^3 (\varphi ^2-8) w+r (-2 (J^2+r^2 v) (r (\varphi ^2-8) w
J''+2 J (\varphi ^2-8) w'-4 J \varphi  w \varphi ')+2 J r (\varphi
^2-8)  (J')^2w+\nonumber\\& &J' ((\varphi ^2-8) (2 r w' (J^2+r^2
v)-4 J^2 w+r^3 w v'+2 r^2 v w)-4 r \varphi  w \varphi ' (J^2+r^2
v))-2 J r^2 (\varphi ^2-8) w v')=0.
\end{eqnarray}

Varying the total action (1) with respect to the scalar field $\varphi$
\begin{eqnarray}
\frac{\partial(\sqrt{-g} L)}{\partial
    \varphi}-\partial_{\rho}\frac{\partial(\sqrt{-g}
    L)}{\partial(\partial_{\rho}\varphi)}=0,
\end{eqnarray}
leads to the Klein-Gordon equation
\begin{eqnarray}
& & \sum _{n=1}^m n  \lambda _{n} \varphi ^{n}+ \frac{\alpha  \varphi ^2}{2 (\varphi ^2-8)^2 w^3 (J^2+r^2 v)}
((J^2+r^2 v) (2 \varphi  (\varphi ^2-8) w \varphi ''-16 w (\varphi ')^2)-\nonumber\\
& &\varphi  (\varphi ^2-8) \varphi ' (2 w' (J^2+r^2 v)-2 J w J'-r
w(r v'+2 v)))
+\frac{\varphi }{16 w^3 (J^2+r^2 v)^2} \big[2 (J^2+r^2 v)\nonumber\\
& & (w (8 \varphi '' (J^2+r^2 v)+2 J \varphi  J''+r^2 \varphi  v''+8 r v \varphi ')-2 w' (4 \varphi ' (J^2+r^2 v)+r v \varphi ))+2 r v' (\varphi  w (3J^2\nonumber\\
& &+r^2 v)-r (J^2+r^2 v) (\varphi  w'-4 w \varphi '))+4 J^2 v \varphi  w+4 J J' (-(J^2+r^2 v) (\varphi  w'-4 w \varphi ')-r^2 \varphi  w v'\nonumber\\
& &-2 r v \varphi  w)-\varphi  w (J')^2 (J^2-3 r^2 v)+r^4 \varphi
(-w) (v')^2\big]=0.
\end{eqnarray}

\section{Solutions }

In this section solutions of the Einstein field equations and Klein-Gordon equation in (2+1) dimensional space-time are given by considering self-interacting scalar fields as an external source for torsion of space-time.

The field equations (13)-(17) and Klein-Gordon equation (19) are a set of coupled differential equations. We shall examine a class of solutions to these equations.

\subsection{Solution with $J=0$ and $\alpha=0$}

The line element (2+1) dimensional space-time in a homogeneous and isotropic universe (8) with $J(r)=0$ is considered as
\begin{eqnarray}
ds^2 = -v(r) dt^2+w(r)^2 dr^2+r^2 d\phi^2.
\end{eqnarray}
In this non-rotating case, the metric component $w(r)$ is given by $ w(r)=\frac{1}{\sqrt{v(r)}} $.

From the Einstein field equations (13) and (14), for the zero-torsion, $\alpha=0$, and the non-rotating case ,$J(r)=0$, we can obtain the scalar field
$\varphi(r)$ and the metric component $v(r)$  respectively as
\begin{eqnarray}
\varphi (r)=\sqrt{\frac{A}{r+B}},
\end{eqnarray}
\begin{eqnarray}
v(r) =\frac{8 A}{r \varphi (r)^4-A \varphi (r)^2+8 A}\big[-8
A^2+C-\Lambda  r^2\sum _{n=1}^m \frac{\lambda _{n } \varphi
(r)^{n-4}}{(n-4) (n-2)}-4 A r \sum _{n=1}^m \frac{\lambda _{n }
\varphi (r)^{n-2}}{n-2}\big],
\end{eqnarray}
where $A$, $B$ and $ C$ are arbitrary constants. The scalar field $\varphi(r)$ is in agreement with those derived in Refs. \cite{Ozcelik2,19,20,Ozcelik1}. The scalar field $\varphi(r)$ goes to zero as $r$ goes to infinity.

From Eq. (15) and Eq. (19), when we set  $\alpha=0$ and $J(r)=0$, we can find
\begin{eqnarray}
& &\sum _{n=1}^m \frac{ \lambda _n}{(n-4) (n-2)}(
\frac{A}{r+B})^{\frac{n}{2}} \big[-A^3 B^2 r (n-6) (n-4) (n-2) +8
A^2 B (B+r) (-32 B^2+B(n\nonumber\\& &(3 (n-10) n+80)-112) r+2 (n
((n-8) n+8)+8) r^2)+64 A (B+r)^3 (32 B^2+B n (-3 (n-8)
n\nonumber\\& &-44) r-(n-2)^2 n r^2)+ 512nr(n-4) (n-2)
(B+r)^5\big]+ 2 A C (A B (2 B+3 r)-8(2 B-r) \nonumber\\&
&(B+r)^2)-2 A B \Lambda  r (3 A^2 B+8 A (-6 B^2-6 B r+r^2)+192 B
(B+r)^2).
\end{eqnarray}

From the above equation, we can obtain
\begin{eqnarray}
& &\text{if} \text{ } n\neq 6,\ \text{}\lambda _n=0, \nonumber\\& &
\text{if} \text{ } n = 6, \;\text{ }A=8B, \text{ } \text{and} \text{ }\nonumber\\& & C=32 B^2 (\Lambda -384 \lambda _6).
\end{eqnarray}
The $\varphi(r)^6$ potential in (2 + 1)-dimensional gravity was
previously known \cite{henne,mar3,mad} to have good behavior in yielding exact black hole solutions.

Under the conditions (24) we can rearrange the scalar field $\varphi(r)$ (21) and $v(r)$ (22) as follows
\begin{eqnarray}
& &\varphi (r)=\sqrt{\frac{8B}{r+B}},
\end{eqnarray}
\begin{eqnarray}
v(r)=-\Lambda  r^2+3 B^2 (\Lambda -512 \lambda _6)+\frac{2 B^3
(\Lambda -512 \lambda _6)}{r}.
\end{eqnarray}

The metric component $ v(r) $ can be written as follows
\begin{eqnarray}
v(r)=-\Lambda  r^2+M+\frac{2 M}{3} \frac{B}{r}
\end{eqnarray}
with $ M=3 B^2 (\Lambda -512 \lambda _6) $. It is clear that for $\Lambda<0$ this is the Mart\'{i}nez-Zanelli solution \cite{19,pot}.

The scalar curvature of the (2+1) space-time (20) with $\alpha=0$ and $J=0$ is obtained as follows
\begin{eqnarray}
R=\frac{8 \left(6 \left(\varphi ^2-8\right) \left(\lambda _6 \varphi^6-\Lambda \right)+8 v (\varphi ')^2\right)}{\left(\varphi ^2-8\right)^2}.
\end{eqnarray}
We can see that the Ricci scalar $R$ (28) goes to $6 \Lambda$ in the limits $r$ going to infinity.

\subsection{Solution with $ J\neq 0$ and $\alpha=0$ }

From the Einstein field equations (13)-(17) and the Klein-Gordon equation (19) with
$\alpha = 0$ and $ J \neq 0$, we can obtain
\begin{eqnarray}
v'(r) &=& \frac{-1}{2 r (2 r \varphi  \varphi '+\varphi ^2-8)}\big[8 J \varphi  J'\varphi '+(\varphi ^2-8)( J')^2-16 J^2 (2 w^2 (\Lambda -\rho _6 \varphi ^6)-(\varphi ')^2)\nonumber\\
& &-8 r v (4 r w^2 (\Lambda -\rho _6 \varphi ^6)-\varphi ' (2 r
\varphi '+\varphi )\big],
\end{eqnarray}
\begin{eqnarray}
w'(r)&=&\frac{w}{2 r (\varphi ^2-8) (J^2+r^2 v) (r J'-2 J)}\big[2 r^2J ((\varphi ^2-8) (r v'-(J')^2)-4 r v \varphi  \varphi ')\nonumber\\
& &+r^3 (2 r v (\varphi ^2-8) J''+J' (4 r v \varphi  \varphi '-(\varphi ^2-8) (r v'+2 v)))+2 r J^2 (r (\varphi ^2-8) \nonumber\\
& &J''+2 J' (r \varphi \varphi '+\varphi ^2-8))-4 J^3 (2 r \varphi
\varphi '+\varphi ^2-8)\big],
\end{eqnarray}
\begin{eqnarray}
\varphi''(r) &=& \frac{1}{8 r^2 \varphi w (J^2+r^2 v)}\big[r^2 (4 r v((\varphi ^2-8+2 r \varphi \varphi ') w'+8 r w^3 (\Lambda -\rho _6 \varphi ^6)+2 w \varphi ' \nonumber\\
& &(r \varphi '-\varphi ))-(\varphi ^2-8) w(J')^2)+4 r J(\varphi ^2-8) wJ'+4 J^2 (8 r^2 w^3 (\Lambda -\rho _6 \varphi ^6)\nonumber\\
& &-w(-2 r^2 \varphi '^2+2 r \varphi  \varphi '+\varphi ^2-8)+r (2
r \varphi \varphi '+\varphi ^2-8) w')\big],
\end{eqnarray}
\begin{eqnarray}
J''(r)&=&\frac{1}{2 r (\varphi ^2-8) (J^2+r^2 v) (2 r \varphi  \varphi'+\varphi ^2-8)}\big[-2 J^3 r (4 (\varphi ')^2 (2 r \varphi  \varphi '-3 \varphi ^2+24)+ \nonumber\\
& &\Lambda  \varphi  w^2 (6 r \varphi ^2 \varphi'-64 r \varphi '+3 \varphi ^3-24 \varphi ))+ 8 \rho _6 r \varphi ^6 w^2 (J' r-2 J) (J^2+r^2 v) (2 r \varphi  \varphi ' \nonumber\\
& &-3 \varphi ^2+24)+J^2 J' (32 (3 r^2 (\varphi ')^2+4)+ \varphi ^4 (3 \Lambda  r^2 w^2+2)+2 r \varphi ^3 \varphi ' (3 \Lambda  r^2 w^2+10) \nonumber\\
& &-4 \varphi ^2 (3 r^2 ((\varphi ')^2+2 \Lambda  w^2)+8)+8 r \varphi  \varphi ' (r^2 ((\varphi ')^2-8 \Lambda  w^2)-20))-2 J r^2 (2 \varphi  \nonumber\\
& & (\varphi ^2-8) \varphi ' (3 (J')^2-4 v)+2 r^2 v \varphi  \varphi ' (4 (\varphi ')^2+\Lambda  (3 \varphi ^2-32) w^2)+3 r v (\varphi ^2-8)  \nonumber\\
& &(\Lambda  \varphi ^2 w^2-4 (\varphi ')^2))+J' r^2 (2 (J')^2 r \varphi   (\varphi ^2-8) \varphi '+v (32 (3 r^2 (\varphi ')^2+4)+\varphi ^4  \nonumber\\
& &(3 \Lambda  r^2 w^2+2)+2 r \varphi ^3 \varphi ' (3 \Lambda  r^2 w^2-2) -4 \varphi ^2 (3 r^2 ((\varphi ')^2+2 \Lambda  w^2)+8)+8 r \varphi  \varphi'  \nonumber\\
& &(r^2 ((\varphi ')^2-8 \Lambda  w^2)+4)))\big].
\end{eqnarray}

The scalar curvature of the (2+1) space-time with $\alpha=0$ and $J(r)\neq0$ is obtained as follows
\begin{eqnarray}
R=\frac{8 \left(8 (\varphi')^2+6 \left(\varphi ^2-8\right) w^2 \left(\rho _6 \varphi ^6-\Lambda \right)\right)}{\left(\varphi ^2-8\right)^2 w^2}.
\end{eqnarray}

The search for exact solutions of the coupled system of differential equations $v'(r)$ (29), $w'(r)$ (30), $\varphi''(r)$ (31), and $J''(r)$ (32) is a very hard job.
We can not solve these equations in analytical forms. These equations are solved numerically with the 4th order Runge-Kutta method.

The plots of the scalar field $\varphi(r)$, the angular momentum $J(r)$, the metric components $v(r)$ and $w(r)$, and the Ricci scalar $R$ with $\alpha=0$ and the angular momentum $J(r)\neq0$ are given in respectively Fig. 1, Fig. 2, Fig. 3, Fig. 4 and Fig. 5. We set $\Lambda=10^{-8}$, $\lambda _6=10^{-11}$, $\varphi(1)=10$, $\varphi'(1)=-2.39583$, $v(1)=10^2$, $w(1)=10^{-1}$, $J(1)=10^{-2}$ and $J'(1)=10^{-4}$.
\begin{figure}[!hbt]
\begin{center}$
\begin{array}{cc}
\scalebox{0.7}{\includegraphics{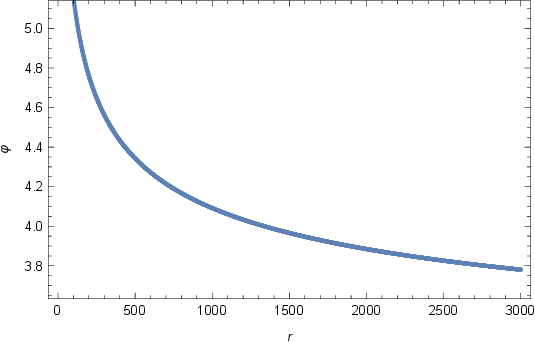}}
\end{array}$
\end{center}
\caption{The scalar field $\varphi(r)$ with $J(r)\neq0$, $\alpha=0$ is plotted by
Runge-Kutta method with respect to $r$
}
\end{figure}
\begin{figure}[!hbt]
\begin{center}$
\begin{array}{cc}
\scalebox{0.72}{\includegraphics{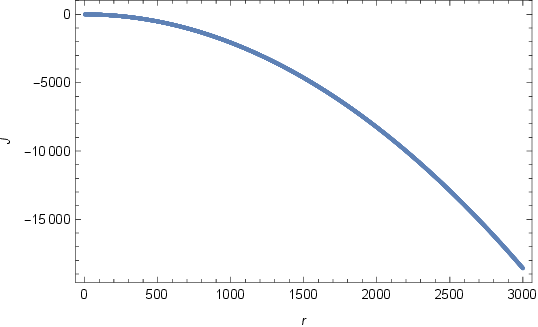}}
\end{array}$
\end{center}
\caption{The angular momentum $J(r)$ with $\alpha=0$ is plotted by
Runge-Kutta method with respect to $r$}
\end{figure}

\begin{figure}[!hbt]
\begin{center}$
\begin{array}{cc}
\scalebox{0.7}{\includegraphics{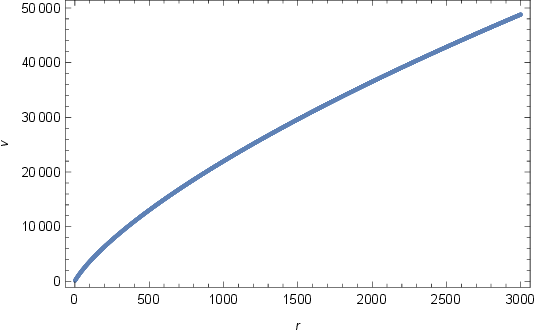}}
\end{array}$
\end{center}
\caption{The metric component $v(r)$ with $J(r)\neq0$, $\alpha=0$ is plotted by
Runge-Kutta method with respect to $r$}
\end{figure}
\begin{figure}[!hbt]
\begin{center}$
\begin{array}{cc}
\scalebox{0.7}{\includegraphics{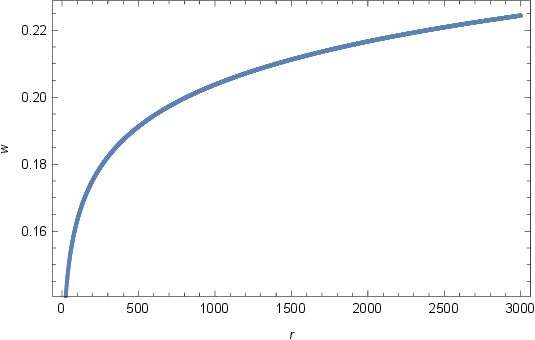}}
\end{array}$
\end{center}
\caption{The metric component $w(r)$
with $J(r)\neq0$, $\alpha=0$ is plotted by
Runge-Kutta method with respect to $r$ }
\end{figure}
\begin{figure}[!hbt]
\begin{center}$
\begin{array}{cc}
\scalebox{0.8}{\includegraphics{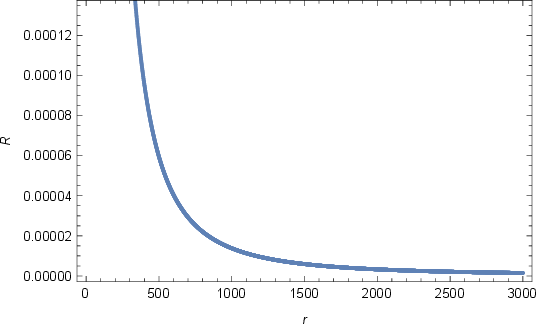}}
\end{array}$
\end{center}
\caption{The Ricci scalar $R$ with $J(r)\neq0$, $\alpha=0$ is plotted by
Runge-Kutta method with respect to $r$}
\end{figure}

\subsection{Solution with $J \neq 0$ and $ \alpha = 1$ }

We consider the metric (8) for a massless self-interacting scalar field non-minimally coupled to Einstein gravity with torsion. The Einstein field equations (13)-(17) and the Klein-Gordon equation (19) are reduced to a system of first and second-order differential equations with torsion $\alpha = 1$ and angular momentum $J \neq 0$ as follows
\begin{eqnarray}
v'(r)&=&\frac{1}{2 r (\varphi ^2-8) (2 r \varphi  \varphi
'+\varphi ^2-8)}\big[(\varphi ^2-8) (32 w^2 (J^2+r^2 v) (\Lambda
-\lambda _6 \varphi ^6)-(\varphi ^2-8)\nonumber\\& &(J')^2)-32
(\varphi ^2-4) (\varphi ')^2 (J^2+r^2 v)-8 \varphi (\varphi ^2-8)
\varphi '(J J'+r v)\big],
\end{eqnarray}
\begin{eqnarray}
w'(r)&=&\frac{w}{4 r (\varphi ^2\!-\!8)^2 (J^2+r^2 v) (2 r \varphi
\varphi ' +\varphi ^2\!-\!8)} \big[r^2 (\varphi ^2-8)^2 (J')^2 (4
r \varphi  \varphi '+\varphi ^2-8)-\nonumber\\& & 4 r J (\varphi
^2-8)^2 J' (4 r \varphi \varphi '+\varphi ^2-8)+4 J^2 (-32 r^3
\varphi  (\varphi ')^3-4 r^2 (\varphi ^4-12 \varphi
^2+32)\nonumber\\& &
 (\varphi ')^2+4 r^2 w^2 (2 r \varphi  \varphi ' (32 \Lambda -7 \Lambda
 \varphi ^2+4\rho _6 \varphi ^8-8 \rho _6 \varphi ^6)+(\varphi ^2-8)
 (16 \Lambda -5 \Lambda\varphi ^2   \nonumber\\& &+2 \rho _6 \varphi ^8+8 \rho _6 \varphi ^6))
 +4 r \varphi  (\varphi ^2-8)^2
\varphi ' +(\varphi ^2-8)^3)+16 r^4 v (w^2 (2 r \varphi  \varphi
'(32 \Lambda -\nonumber\\& &7 \Lambda  \varphi ^2+4 \rho _6
\varphi ^8-8 \rho _6 \varphi ^6)+(\varphi ^2-8) (16 \Lambda -5
\Lambda  \varphi ^2+2 \rho _6 \varphi ^8+8 \rho _6 \varphi
^6))-\nonumber\\& &(\varphi ')^2 (8 r \varphi  \varphi '+\varphi
^4-12 \varphi ^2+32))\big],
\end{eqnarray}
\begin{eqnarray}
\varphi ''(r)&=&\frac{1}{2 r (\varphi ^2-8)^2 (J^2+r^2
v)}\big[\varphi ' ((\varphi ^2-8)^2  (r^2 ((J')^2-2 v)-4 r J J'+2
J^2)+2 r \varphi \nonumber\\& &  (\varphi ^2-8) (J^2+r^2 v)
\varphi '-32 r^2 (J^2+r^2 v)  (\varphi ')^2)+4 w^2 (2 r \varphi '
(32 \Lambda -7 \Lambda  \varphi ^2+4 \rho _6 \varphi ^8
\nonumber\\& &-8 \rho _6 \varphi ^6)-3 \varphi (\varphi ^2-8)
(\Lambda -8 \rho _6 \varphi ^4))\big],
\end{eqnarray}
\begin{eqnarray}
J''(r)&=&\frac{1}{r (\varphi ^2-8)^2 (J^2+r^2 v) (2 r \varphi
\varphi '+\varphi ^2-8)} \big[4 J^2 (-32 r^3 \varphi  (\varphi
')^3-4 (\varphi ^4-12 \varphi ^2+32)
 \nonumber\\& &r^2(\varphi ')^2+4 r^2 w^2 ((\varphi ^2-8) (-5 \Lambda  \varphi ^2+
 16\Lambda +2 \rho _6 \varphi ^8+8 \rho _6 \varphi ^6)+2 r \varphi  \varphi '
 (-7 \Lambda  \varphi ^2+\nonumber\\& & 32 \Lambda+4 \rho _6 \varphi ^8-8 \rho _6 \varphi ^6))
 +4 r (\varphi ^2-8)^2\varphi  \varphi '+(\varphi ^2-8)^3)+r^2 (\varphi ^2-8)^2 (J')^2
  (4 r\nonumber\\& & \varphi  \varphi '+\varphi ^2-8)-4  r (\varphi ^2-8)^2
  (4 r \varphi  \varphi '+\varphi ^2-8) J J'+16 r^4 v (w^2 ((\varphi ^2-8)
  (-5 \Lambda \nonumber\\& & \varphi ^2+16 \Lambda +2 \rho _6 \varphi ^8+8 \rho _6 \varphi ^6)
  +2 r \varphi  \varphi ' (-7 \Lambda  \varphi ^2+32 \Lambda +4 \rho _6 \varphi ^8-
  8 \rho _6 \varphi ^6))\nonumber\\& &-(\varphi ')^2 (8 r \varphi  \varphi '+\varphi ^4-12 \varphi ^2+32))\big].
\end{eqnarray}

The curvature scalar $R$ with $\alpha=1$ and the angular momentum $J(r)\neq0$ can be present as
\begin{eqnarray}
R=-\frac{8 \left((\varphi')^2+6 \Lambda  w^2-6 \lambda _6 w^2\varphi ^6 \right)}{\left(\varphi ^2-8\right) w^2}.
\end{eqnarray}
In the absence of the coupling constant $\lambda _6$, the curvature scalar of space-time with torsion is reduced to $R(r)$ (Eq. (50)) in Ref. \cite{Ozcelik2}.

The equations $v'(r)$ (34), $w'(r)$ (35), $\varphi''(r)$ (36), and $J''(r)$ (37) can not be solved in analytical forms. We will give only the numerical solutions using the methods of 4th order Runge-Kutta.

The plot of the scalar field $\varphi(r)$, the angular momentum $J(r)$, the metric components $v(r)$ and $w(r)$, and the Ricci scalar with torsion are given  respectively in  Fig. 6, Fig. 7, Fig. 8, Fig. 9 and Fig. 10. We set $\Lambda=10^{-8}$, $\lambda _6=10^{-11}$, $\varphi(1)=10$, $\varphi'(1)=-2.39583$, $v(1)=10^2$, $w(1)=10^{-1}$, $J(1)=10^{-2}$ and $J'(1)=10^{-4}$.
\begin{figure}[!hbt]
\begin{center}$
\begin{array}{cc}
\scalebox{0.7}{\includegraphics{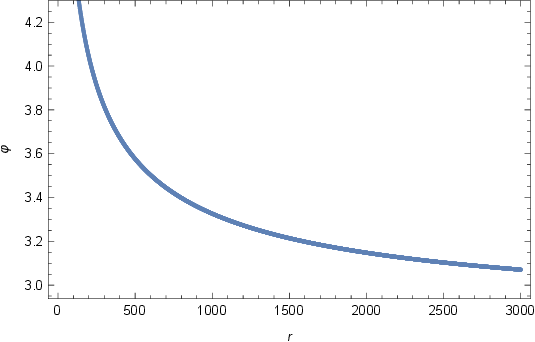}}
\end{array}$
\end{center}
\caption{The scalar field $\varphi(r)$ with $J(r)\neq0$, $\alpha=1$ is plotted by
Runge-Kutta method with respect to $r$}
\end{figure}
\begin{figure}[!hbt]
\begin{center}$
\begin{array}{cc}
\scalebox{0.7}{\includegraphics{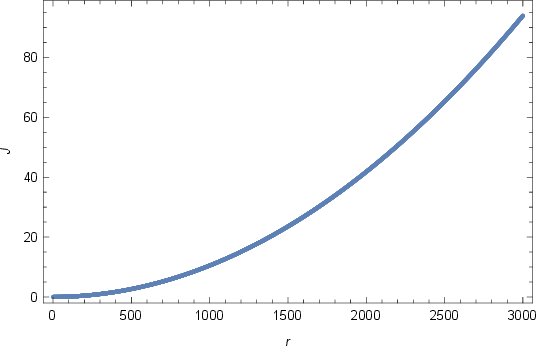}}
\end{array}$
\end{center}
\caption{The angular momentum $J(r)$ with $\alpha=1$ is plotted by
Runge-Kutta method with respect to $r$}
\end{figure}
\begin{figure}[!hbt]
\begin{center}$
\begin{array}{cc}
\scalebox{0.7}{\includegraphics{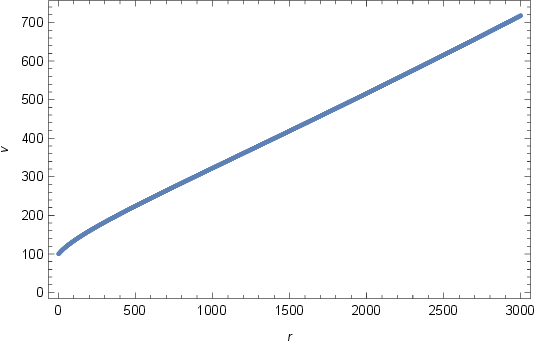}}
\end{array}$
\end{center}
\caption{The metric component $v(r)$ with $J(r)\neq0$, $\alpha=1$ is plotted by
Runge-Kutta method with respect to $r$ }
\end{figure}
\begin{figure}[!hbt]
\begin{center}$
\begin{array}{cc}
\scalebox{0.7}{\includegraphics{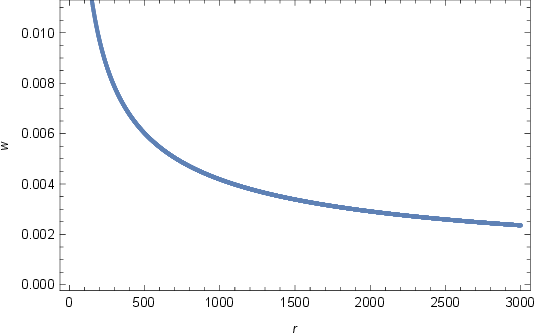}}
\end{array}$
\end{center}
\caption{The metric component
$w(r)$ with $J(r)\neq0$, $\alpha=1$ is plotted by
Runge-Kutta method with respect to $r$ }
\end{figure}
\begin{figure}[!hbt]
\begin{center}$
\begin{array}{cc}
\scalebox{0.7}{\includegraphics{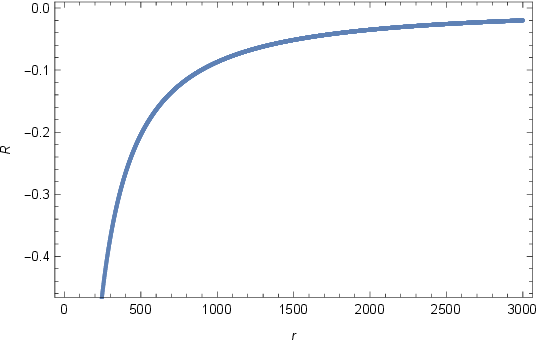}}
\end{array}$
\end{center}
\caption{The Ricci scalar $R$ with $J(r)\neq0$, $\alpha=1$ is plotted by
Runge-Kutta method with respect to $r$ }
\end{figure}

We see that the Figs. 6-10 for $J\neq0$, $\alpha=1$ case are distinctly different from the $J\neq0$, $\alpha=0$ case, given in Figs. 1-5. We can deduce that the torsion has an effect on the scalar field $\varphi(r)$, the angular momentum $J(r)$, the metric components $v(r)$, $w(r)$, and the curvature scalar $R(r)$.

\section{The equations of motion}

To explore the (2+1) dimensions of Einstein gravity with torsion induced by the self-interacting scalar fields further, let us consider the motion of freely moving both massive and massless particles.

Test particles move along geodesics of (2+1) space-time; the geodesics equations being derivable from the Lagrangian

\begin{eqnarray}
L=-\frac{\varepsilon ^2}{2}=\frac{1}{2}g_{\mu \nu }  \dot{x}^{\mu }\dot {x}^{\nu },
\end{eqnarray}
where $\varepsilon=1$ corresponds to time-like geodesics, $\varepsilon=0$ corresponds to null geodesics. $\tau$ is the proper time for massive particles $\varepsilon=1$ and affine parameter for massless particles $\varepsilon=0$. Here an overdot denotes the partial derivative with respect to an affine parameter $\tau$.

The Euler-Lagrangian equations are
\begin{eqnarray}
\frac{d}{d\tau }\frac{\partial L}{\partial \dot{x^{\mu}}}-\frac{\partial L}{\partial x^{\mu}}=0.
\end{eqnarray}
Using the Lagrangian (39) and the Euler-Lagrangian equation (40), the constants of the motion corresponding to the ignorable coordinates $t$ and $\phi$are  calculated for the metric (8)

\begin{eqnarray}
J \dot{\phi }-\dot{t} v=-E,
\end{eqnarray}
\begin{eqnarray}
J \dot{t}+r^2 \dot{\phi }=L.
\end{eqnarray}
From the Lagrangian (39) we can obtain the constant of motion corresponding to the conservation of rest mass
$\varepsilon$

\begin{eqnarray}
-v \dot t^2+2 J \dot{t} \dot{\phi }+ w^2 \dot r^2+r^2 \dot{\phi }^2=-\varepsilon^2.
\end{eqnarray}
The Eqs. (41), (42), and (43) can be solved simultaneously for $\dot t$, $\dot \phi$, and $\dot r$ to give
\begin{eqnarray}
\dot{t}=\frac{J L+E r^2}{J^2+r^2 v}
\end{eqnarray}
\begin{eqnarray}
\dot{\phi }=\frac{-E J+L v}{J^2+r^2 v}
\end{eqnarray}
\begin{eqnarray}
\dot{r}^2=\frac{E ^2 r^2+2 E J L-L^2 v-\varepsilon^2 (J^2+r^2 v)}{(J^2+r^2 v)w^2}
\end{eqnarray}

Using Eq. (46), we can find the effective potential $\mathbb{V}(r)$ for radial motion as follows
\begin{eqnarray}
\mathbb{V}(r)=\frac{-J L\mp \sqrt{(J^2+r^2 v) (L^2+\varepsilon^2 r^2 )}}{r^2}.
\end{eqnarray}
The effective potential determines allowed regions of motion.

\subsection{\label{}  The effective potential with $J=0$ and $\alpha=0$}

Substituting the metric component $v(r)$ (27) and $J=0$ into Eq. (47) we find the effective potential as
\begin{eqnarray}
\mathbb{V}(r)=\frac{\sqrt{(L^2+ \varepsilon^2 r^2 ) (\frac{2 B M}{3 r}+M-\Lambda  r^2)}}{r},
\end{eqnarray}
with $ M=3 B^2 (\Lambda -512 \lambda _6) $.

a) In the case of massless particles $\varepsilon=0$, the results of the numerical analysis of the effective potential $\mathbb{V}(r)$ for several $L=10, 1000, 5000$ and $10000$ are plotted for in Fig. 11.

In this case, the effective potential for the radial motion with $L=0$ is $\mathbb{V}(r)=0$. Therefore the massless particles behaves as free particles. We set $\Lambda =10^{-8}$, $\lambda_6=10^{-11}$ and $B=2.10^4$.

The real positive root of the potential for $\epsilon=0$ coincides with the horizon value $r_h=29205.3$. The effective potential for massless particles assumes imaginary values $r>r_h$ for $L \neq 0$.

\begin{figure}[!hbt]
\begin{center}$
\begin{array}{cc}
\scalebox{0.8}{\includegraphics{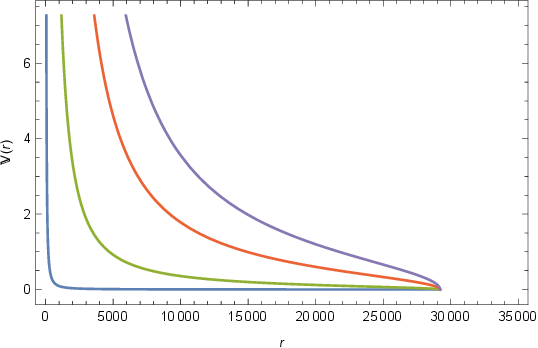}}
\end{array}$
\end{center}
\caption{The effective potential $\mathbb{V}(r)$ with $J(r)=0$, $\alpha=0$ for massless particles $\varepsilon =0$. We set $\Lambda =10^{-8}$, $\lambda_6=10^{-11}$ and $B=2.10^4$. The figure shows the effective potential for the four different values of the angular momentum $L=10, 1000, 5000$, and $10000$ from the bottom up}
\end{figure}

b) In the case of massive particles $\varepsilon=1$, the results of the numerical analysis of the effective potential $\mathbb{V}(r)$ (48) for several $L=10, 1000, 5000$ and $10000$ are plotted  in Fig. 12.

The real positive root of the effective potential for  $\epsilon=1$ coincides with the horizon value $r_h=29205.3$. The effective potential for massive particles assumes imaginary values $r>r_h$ for different values of $L$.

\begin{figure}[!hbt]
\begin{center}$
\begin{array}{cc}
\scalebox{0.8}{\includegraphics{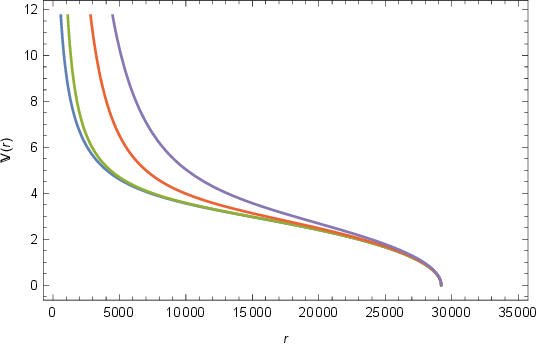}}
\end{array}$
\end{center}
\caption{The effective potential $\mathbb{V}(r)$ with $J(r)=0$, $\alpha=0$ for massive particles $\varepsilon =1$. We set $\Lambda =10^{-8}$, $\lambda_6=10^{-11}$ and $B=2.10^4$. The figure shows the effective potential for the four different values of the angular momentum $L=10, 1000, 5000$, and $10000$ from the bottom up}
\end{figure}

\subsubsection{Trajectory of the particles for $J=0$ and $\alpha=0$}

a) In the case of massless particles $\varepsilon=0$, the solutions of the equations of motion (44), (45) and (46) are given by

\begin{eqnarray}
r(\tau )=\frac{1}{2} \tau  \sqrt{E^2+\Lambda  L^2} + D_1,
\end{eqnarray}
\begin{eqnarray}
\phi (\tau )&=\frac{2 L \left(2 D_1\sqrt{E^2+\Lambda  L^2}-\tau
\left(E^2+\Lambda  L^2\right)\right)}{\left(E^2+\Lambda L^2\right)
\left(\tau ^2 \left(E^2+\Lambda  L^2\right)-4 D_1{}^2\right)}+D_2,
\end{eqnarray}
\begin{eqnarray}
t (\tau)=\frac{2 E\left(2 C_3 \sqrt{E^2+\Lambda  L^2}-\tau
\left(E^2+\Lambda  L^2\right)\right)}{\Lambda  \left(E^2+\Lambda
L^2\right) \left(\tau ^2 \left(E^2+\Lambda  L^2\right)-4
D_1{}^2\right)}+D_3,
\end{eqnarray}
Here $D_1$, $D_2$ and $D_3$ are constants.

We have plotted $r(\tau)$ (49), $\phi(\tau)$ (50), and $t(\tau)$ (51) for massless particles $\varepsilon=0$ as a function of $\tau$ in Figs. 13, 14, and 15 respectively.

\begin{figure}[!hbt]
\begin{center}$
\begin{array}{cc}
\scalebox{0.7}{\includegraphics{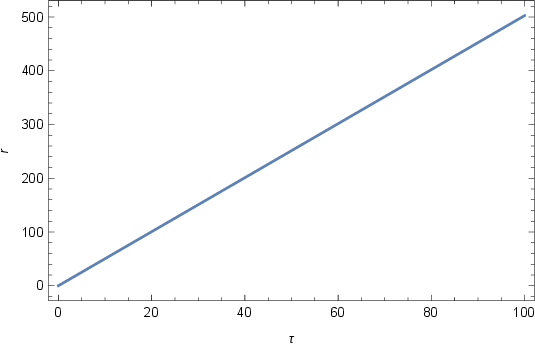}}
\end{array}$
\end{center}
\caption{$r(\tau)$ with $J(r)=0$, $\alpha=0$ for massless particles $\varepsilon=0$ is plotted with respect to $\tau$. We set $\Lambda=10^{-4}$, $E=10$, $L=100$ and $D_1=0$}
\end{figure}
\begin{figure}[!hbt]
\begin{center}$
\begin{array}{cc}
\scalebox{0.7}{\includegraphics{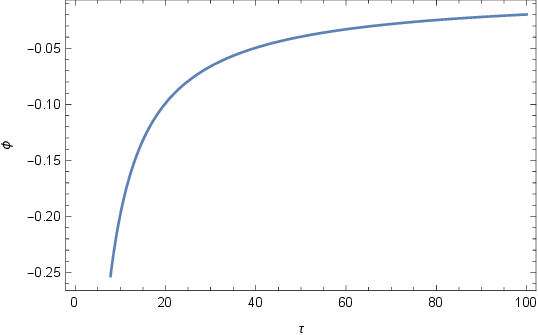}}
\end{array}$
\end{center}
\caption{$\phi(\tau)$ with $J(r)=0$, $\alpha=0$ for massless particles $\varepsilon=0$ is plotted with respect to $\tau$. We set $\Lambda=10^{-4}$, $E=10$, $L=100$, $D_1=0$, and $D_2=0$}
\end{figure}

\begin{figure}[!hbt]
\begin{center}$
\begin{array}{cc}
\scalebox{0.7}{\includegraphics{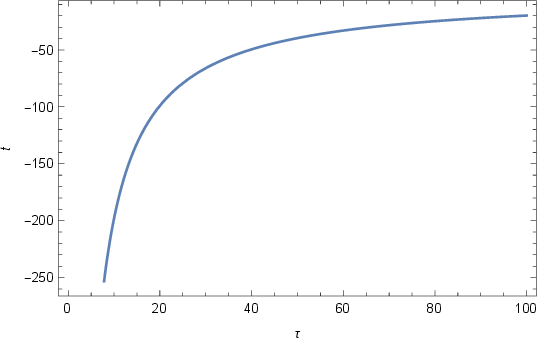}}
\end{array}$
\end{center}
\caption{$t(\tau)$ with $J(r)=0$, $\alpha=0$ for massless particles $\varepsilon=0$ is plotted with respect to $\tau$. We set $\Lambda=10^{-4}$, $E=10$, $L=100$, $D_1=0$ and $D_3=0$}
\end{figure}

\begin{figure}[!hbt]
\begin{center}$
\begin{array}{cc}
\scalebox{0.8}{\includegraphics{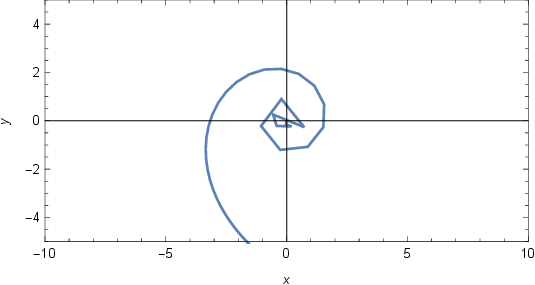}}&\\
\scalebox{0.8}{\includegraphics{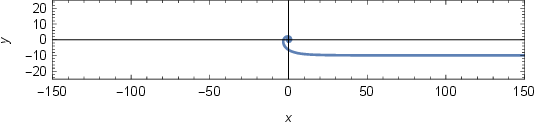}}
\end{array}$
\end{center}
\caption{The equatorial orbits with $J(r)=0$, $\alpha=0$ for massless particles $\varepsilon=0$ are plotted. Here $x=r cos \phi$ and $y= r sin \phi$. We set $\Lambda=10^{-4}$, $E=10$, $L=100$, $D_1=0$ and $D_2=0$ and $D_3=0$  }
\end{figure}

In Fig. 16 the equatorial orbits for massless particles $\varepsilon=0$ are shown for $\Lambda=10^{-4}$, $E=10$ and $L=100$.

In the upper part of Fig. 16, the equatorial orbit is plotted in the region $-10 \leq x \leq10$. In the lower part of Fig. 16
the equatorial orbit is also plotted in the region $-150 \leq x \leq 150$.

b) In the case of massive particles $\varepsilon=1$, the solutions of the equations of motion (44), (45) and (46) are given by
\begin{eqnarray}
r(\tau)=\frac{\sqrt{E^2+\Lambda  L^2}}{\sqrt{2}\sqrt{\varepsilon }
 \sqrt{\Lambda } }\sin \left(\frac{2 D_4 \sqrt{\varepsilon } \sqrt{\Lambda }
 +\tau}{\sqrt{2}}\right)\left| \pm\left(\cos \left(\frac{2 D_4 \varepsilon
  \sqrt{\Lambda }+\tau}{\sqrt{2}}\right)\right)\right|,
\end{eqnarray}
\begin{eqnarray}
\phi(\tau)=-\frac{\sqrt{2}\sqrt{\varepsilon } \sqrt{\Lambda } L
\cot \left(\frac{2 D_4 \sqrt{\Lambda } \sqrt{\varepsilon
}+\tau}{\sqrt{2}}\right)}{E^2+\Lambda  L^2}+D_5,
\end{eqnarray}

\begin{eqnarray}
t(\tau)=-\frac{\sqrt{2} E \varepsilon  \cot \left(\frac{2 D_4 \sqrt{\Lambda } \sqrt{\varepsilon }+\tau} {\sqrt{2}}\right)}{E^2+\Lambda  L^2}+D_6.
\end{eqnarray}
Here $D_4$, $D_5$ and $D_6$ are constants.

We have plotted $r(\tau)$ (52) as a function of $\tau$ in Fig. 17. The trajectory of the massive particles exhibits oscillatory motion which clearly differs from the trajectory for massless particles in Fig. 13. We have also plotted $\phi(\tau)$ (53) and $t(\tau)$ (54) as a function of $\tau$ in Figs. 18 and 19 respectively.

We set $\Lambda=10^{-4}$, $E=10$, $L=100$, $D_4=0$, $D_5=0$ and $D_6=0$.
\begin{figure}[!hbt]
\begin{center}$
\begin{array}{cc}
\scalebox{0.7}{\includegraphics{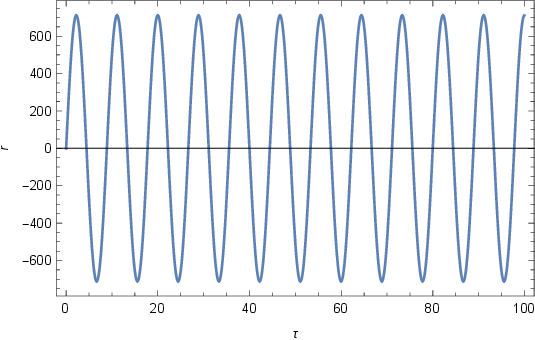}}
\end{array}$
\end{center}
\caption{$r(\tau)$ with $J(r)=0$, $\alpha=0$ for massive particles $\varepsilon=1$ is plotted with respect to $\tau$. We set $\Lambda=10^{-4}$, $E=10$, $L=100$, $D_4=0$}
\end{figure}
\begin{figure}[!hbt]
\begin{center}$
\begin{array}{cc}
\scalebox{0.7}{\includegraphics{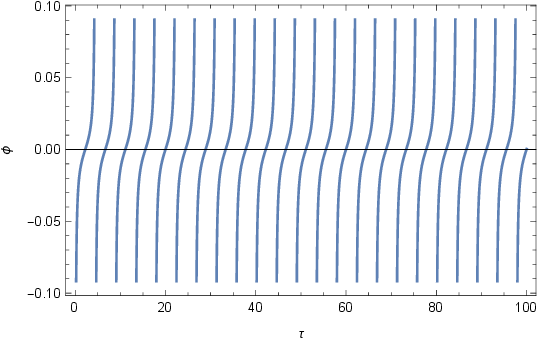}}
\end{array}$
\end{center}
\caption{$\phi(\tau)$ with $J(r)=0$, $\alpha=0$ for massive particles $\varepsilon=1$ is plotted with respect to $\tau$. We set $\Lambda=10^{-4}$, $E=10$, $L=100$, $D_4=0$, $D_5=0$}
\end{figure}
\begin{figure}[!hbt]
\begin{center}$
\begin{array}{cc}
\scalebox{0.7}{\includegraphics{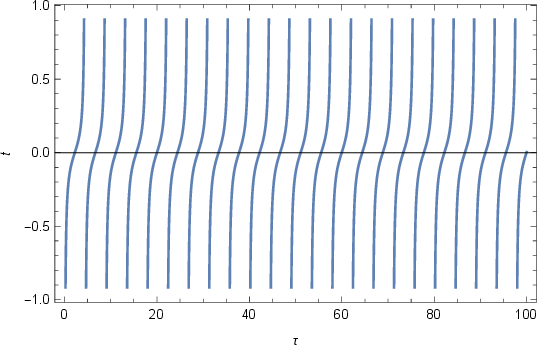}}
\end{array}$
\end{center}
\caption{$t(\tau)$ with $J(r)=0$, $\alpha=0$ for massive particles $\varepsilon=1$ is plotted with respect to $\tau$. We set $\Lambda=10^{-4}$, $E=10$, $L=100$, $D_4=0$, $D_6=0$}
\end{figure}

In Fig. 20 the equatorial orbits for massive particles $\varepsilon=1$ are shown for $\Lambda=10^{-4}$, $E=10$ and $L=100$.
\begin{figure}[!hbt]
\begin{center}$
\begin{array}{cc}
\scalebox{0.8}{\includegraphics{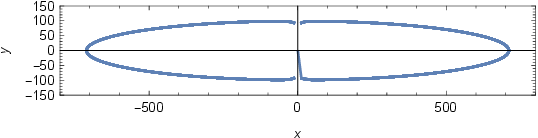}}
\end{array}$
\end{center}
\caption{The equatorial orbits with $J(r)=0$, $\alpha=0$ for massive particles $\varepsilon=1$ is plotted. Here $x=r cos \phi$ and $y= r sin \phi$. We set $\Lambda=10^{-4}$, $E=10$, $L=100$, $D_4=0$, $D_5=0$, $D_6=0$  }
\end{figure}

\subsection{\label{}The effective potential for $ J \neq 0 $ and $ \alpha = 0 $}

In this section, we examine the effect of the non-zero angular momentum $ J \neq 0$ on the radial motion of massless particles and massive particles in the metric (8), respectively.

a) In the case of massless particles $\varepsilon=0$, the effective potential $\mathbb{V}(r)$ (47) for radial motion is plotted by Runge-Kutta method with respect to $r$ in Fig. 21. We set $\Lambda=10^{-8}$,
$\lambda _6=10^{-11}$, $\varphi(1)=10$, $\varphi'(1)=-2.39583$,
$v(1)=10^2$, $w(1)=10^{-1}$, $J(1)=10^{-2}$ and $J'(1)=10^{-4}$.
\begin{figure}[!hbt]
\begin{center}$
\begin{array}{cc}
\scalebox{0.8}{\includegraphics{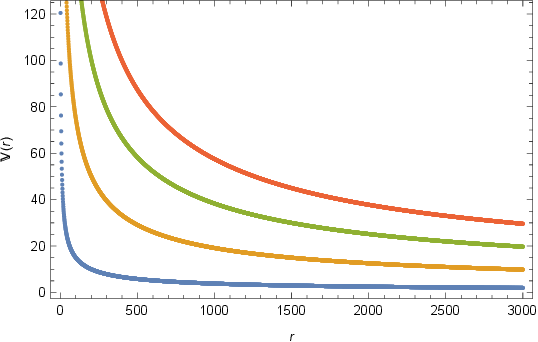}}
\end{array}$
\end{center}
\caption{The plots of the effective potential  $\mathbb{V}(r)$ with $J(r)\neq 0$, $\alpha=0$ for massless particles $\varepsilon =0$ are plotted by Runge-Kutta method with respect to $r$. The figure shows the effective potential for the different values of the angular momentum $L=10, 50, 100$ and $150$ from the bottom up }
\end{figure}

Fig. 21 shows that circular orbits do not exist in this space-time with zero torsion $\alpha=0$.

b) In the case of massive particles $\varepsilon=1$, we plot the effective potential (47) in Fig. 22. We set $\Lambda=10^{-8}$,
$\lambda _6=10^{-11}$, $\varphi(1)=10$, $\varphi'(1)=-2.39583$,
$v(1)=10^2$, $w(1)=10^{-1}$, $J(1)=10^{-2}$ and $J'(1)=10^{-4}$. Fig. 22 shows that there exist marginally stable circular orbits in a (2+1) Einstein gravity with zero-torsion.
\begin{figure}[!hbt]
\begin{center}$
\begin{array}{cc}
\scalebox{0.8}{\includegraphics{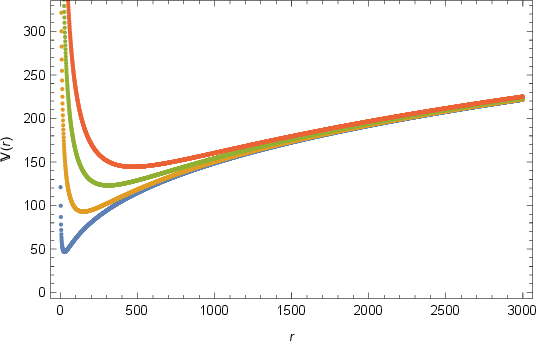}}
\end{array}$
\end{center}
\caption{The plots of the effective potential $\mathbb{V}(r)$ with $J(r)\neq 0$, $\alpha=0$ for massive particles $\varepsilon =1$ are plotted by Runge-Kutta method with respect to $r$. The figure shows the effective potential for the four different curves correspond to the angular momentum $L=10, 50, 100$ and $150$ from the bottom up }
\end{figure}

\subsection{\label{}The effective potential for $ J \neq 0 $ and $\alpha = 1$ }

We study the effective potential $\mathbb{V}(r)$ (47) for radial motion in a (2+1) Einstein gravity with torsion induced by the non-minimally coupled real scalar field interacting with itself for massless particles $\varepsilon=0$ and massive particles $\varepsilon=1$.

a) In the case of massless particles $\varepsilon=0$, the effective potential $\mathbb{V}(r)$ for radial motion is plotted by Runge-Kutta method with respect to $r$ in Fig. 23. We set $\Lambda=10^{-8}$,
$\lambda _6=10^{-11}$, $\varphi(1)=10$, $\varphi'(1)=-2.39583$,
$v(1)=10^2$, $w(1)=10^{-1}$, $J(1)=10^{-2}$, $J'(1)=10^{-4}$. This numerical analysis reveals that there are no stable circular orbits in this space-time with torsion.
\begin{figure}[!hbt]
\begin{center}$
\begin{array}{cc}
\scalebox{0.8}{\includegraphics{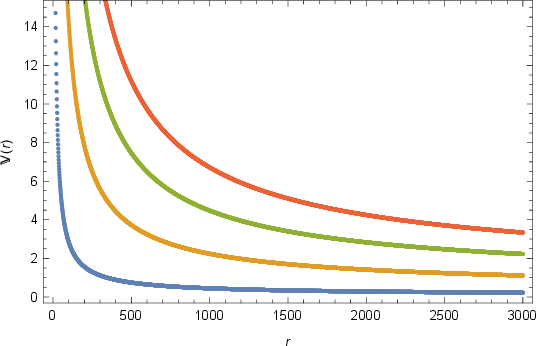}}
\end{array}$
\end{center}
\caption{The plot of the effective potential $\mathbb{V}(r)$ with $J(r)\neq 0$, $\alpha=0$ for massless particles $\varepsilon =0$ is plotted by Runge-Kutta method with respect to $r$. The figure shows the effective potential for the four different curves corresponding to the angular momentum $L=10, 50, 100$, and $150$ from the bottom up }
\end{figure}

b) In the case of massive particles $\varepsilon=1$, the effective potential $\mathbb{V}(r)$ for radial motion is plotted by Runge-Kutta method with respect to $r$ in Fig. 24. We set $\Lambda=10^{-8}$,
$\lambda _6=10^{-11}$, $\varphi(1)=10$, $\varphi'(1)=-2.39583$.
$v(1)=10^2$, $w(1)=10^{-1}$, $J(1)=10^{-2}$, $J'(1)=10^{-4}$.
\begin{figure}[!hbt]
\begin{center}$
\begin{array}{cc}
\scalebox{0.8}{\includegraphics{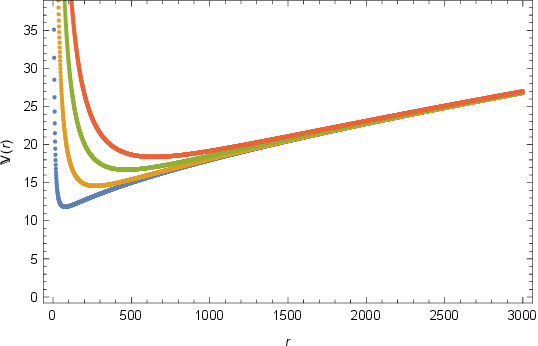}}
\end{array}$
\end{center}
\caption{The plot of the effective potential $\mathbb{V}(r)$ with $J(r)\neq 0$, $\alpha=0$ for massive particles $\varepsilon =1$ is plotted by Runge-Kutta method with respect to $r$. The figure shows the four different curves correspond to the angular momentum $L=10, 50, 100$ and $150$ from the bottom up}
\end{figure}

We show that there exist stable circular orbits in (2+1) Einstein gravity induced by the self-interacting scalar field. The radius of stable orbits increases with increasing $L$ for all positive values of $L$. We can see that the depth of the effective potential well decreases with increasing angular momentum $L$.

 \section{\label{} Conclusions}

In this work, we have studied self-interacting scalar fields in (2+1) dimensions  Einstein gravity with torsion $\alpha=1$ in the presence of a cosmological constant. We have obtained the field equations with a self-interaction potential by a variational principle. We have then investigated non-rotating $J\neq0$ and rotating $J=0$, circularly symmetric solutions. We have obtained analytical solutions of field equations with $J(r)=0$ and $\alpha=0$. Using the Runge-Kutta method, we have given the numerical solutions of these equations with $J(r)\neq0$ for the case of $\alpha=0$ and the case of $\alpha=1$ respectively. In the case of $J\neq0$, $\alpha=1$, the scalar field $\varphi(r)$ goes to zero faster than in the case of $J\neq0$, $\alpha=0$.

We can conclude that the torsion has an effect on the scalar field $\varphi(r)$, the metric components $v(r)$, $w(r)$, the angular momentum $J(r)$, and the Ricci scalar $R$.

The investigation of the motion of test particles may be a very useful tool to study the nature of the gravitational properties of the corresponding space-time metrics.
We perform the numerical calculations with the help of the 4th-order Runge-Kutta method. We have examined the orbits for the effective potential of particles for different cases of $J$, $\alpha$, and different values of $L$ and compared the results to these orbits.

We have concluded that the orbits given by $\epsilon=0$ are not stable but the orbits are stable for $\epsilon=1$.
The depths of the effective potential wells in Fig. 22 are less than those in Fig. 21.

We also have derived first-order equations for particles in the non-rotating $J\neq0$ and torsionless $\alpha=0$ backgrounds and analyzed the properties of some special trajectories.
An explicit numerical study of the effective potential for specific values of the parameters could lead to interesting results.

\section*{ Acknowledgments} We are grateful to Professor Mahmut Horta\c{c}su for reading the manuscript. This work has been supported by Yildiz Technical University Scientific Research Projects Coordination Unit under project number FBA-2021-4686.

\section*{Declarations}
{\bf Data Availability Statement}: This manuscript has no associated data or
the data will not be deposited.

\end{document}